\documentclass[a4paper,11pt]{article}
\usepackage{jinstpub} 
\usepackage{siunitx}
\title{Determining the electron-hole pair creation energy in a 4H-SiC p-n diode via X-ray spectroscopy}

\author[a]{Silas M\"uller}
\author[b,1]{Pascal Wolf}
\author[b,2]{Patrick Ahlburg}
\author[b]{Christian Bespin}
\author[a]{Jochen Dingfelder}
\author[c]{Gr\'egory Grosset}

\affiliation[a]{Physikalisches Institut and Cluster of Excellence ``Color meets Flavor'',
University of Bonn, Nu{\ss}allee 12, 53115 Bonn, Germany}

\affiliation[b]{Physikalisches Institut, University of Bonn,
Nu{\ss}allee 12, 53115 Bonn, Germany}

\affiliation[c]{Ion Beam Services (IBS),
Rue Gaston Imbert prolong\'ee, 13790 Peynier, France}

\affiliation[1]{Now at DECTRIS AG, Baden-D\"attwil, Switzerland}

\affiliation[2]{Now at Black Semiconductor GmbH, Aachen, Germany}
\emailAdd{silas.mueller@uni-bonn.de}
\abstract{This paper reports on a measurement to determine the electron-hole pair creation energy in a 4H-SiC p-n diode. By measuring the spectra of several X-ray and $\gamma$ sources and comparing the response between the silicon-carbide diode and a commercial silicon diode, the electron-hole pair creation energy in 4H-SiC is determined to be \qty{7.87 +- 0.09}{\electronvolt}. This is the first result obtained using X-ray spectra from several sources and including an estimate of the uncertainty. The result is in good agreement with studies of comparable devices.}


\begin{document}
\maketitle
\flushbottom

\section{Introduction}\label{sec:Introduction}
Segmented silicon detectors are often used as precision tracking detectors in high-energy physics experiments as they can be designed for high radiation tolerance, high granularity and fast readout. The properties of silicon for radiation detection have been thoroughly investigated and are widely published. Research for novel detector materials with advantageous properties is ongoing. Silicon carbide (SiC) is a material that exhibits promising characteristics for the use in particle physics experiments. Its wide band gap of \SI{3.23}{\eV} compared to silicon results in lower leakage current, allowing for operation at higher temperatures. However, the average energy for the creation of an electron-hole pair is larger in SiC  and the energy resolution is therefore expected to be worse than for silicon. Furthermore, silicon is widely-used and the processing technique well-established and understood compared to silicon carbide. The large displacement energy of $\SIrange{30}{40}{\eV}$ in SiC compared to $\SIrange{13}{15}{\eV}$ in silicon results in potentially better radiation hardness \cite[p. 3]{SiC}. A high breakdown electric field strength enables operation at high reverse bias voltages, which reduces the transit time of charge carriers and, therefore, the probability of trapping \cite[p. 5]{SiC}. 

In order to design such novel detectors, an understanding of the material properties, e.g. the electron-hole (e-h) pair creation energy, $\omega$, is needed. The literature values range from \SI{7.6}{\eV} \cite{Phlips2005} to \SI{8.6}{\eV} \cite{Lebedev2004}. In this paper, a value for $\omega$ in 4H-SiC is determined by comparing signals from several X-ray and $\gamma$ sources for SiC and Si, scaling the ratio with the known value of $\omega$ for silicon (Section \ref{sec:method}).

\subsection{Method}\label{sec:method}
For the determination of $\omega_{SiC}$, a spectroscopic measurement with various photon energies $E_{\gamma}$ is performed for a Si and a 4H-SiC diode and the results are compared with each other. The mean peak position, $\mu$, of a recorded gamma peak (full-energy peak) is proportional to the number of created e-h pairs: 
\begin{equation}\label{eq}
    \frac{E_{\gamma}}{\omega}=N_{\text{e-h pairs}} \propto \mu.
\end{equation}
The ratio of the peak positions for SiC and Si thus corresponds to the ratio of the e-h pair creation energy in Si ($\omega_{Si}$) and SiC ($\omega_{SiC}$). Consequently, the following equation can be derived:
\begin{equation}
    \frac{\omega_{SiC}}{\omega_{Si}}=\frac{\mu_{Si}}{\mu_{SiC}}.
\end{equation}
With a known value for $\omega_{Si}$, $\omega_{SiC}$ can be calculated:
\begin{equation}\label{eq:Eeh}
    \omega_{SiC}={\omega_{Si}}\cdot\frac{\mu_{Si}}{\mu_{SiC}}.
\end{equation}
In this work, the used electron-hole pair creation energy in silicon is \SI{3.66+-0.03}{\eV} ~\cite{SCHOLZE2000208}.

\section{Diodes and measurement setup}\label{sec:Setup}
\subsection{Diodes}
\begin{figure}[h]
    \centering
    \includegraphics[width=0.495\textwidth]{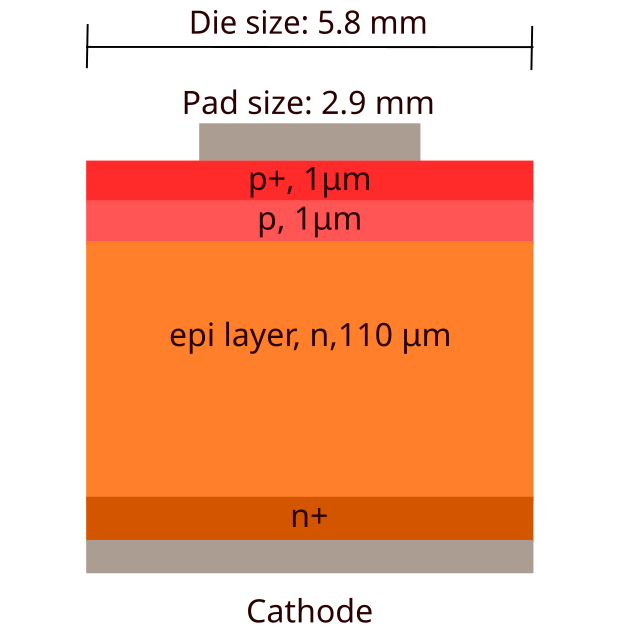}
    \includegraphics[width=0.495\textwidth]{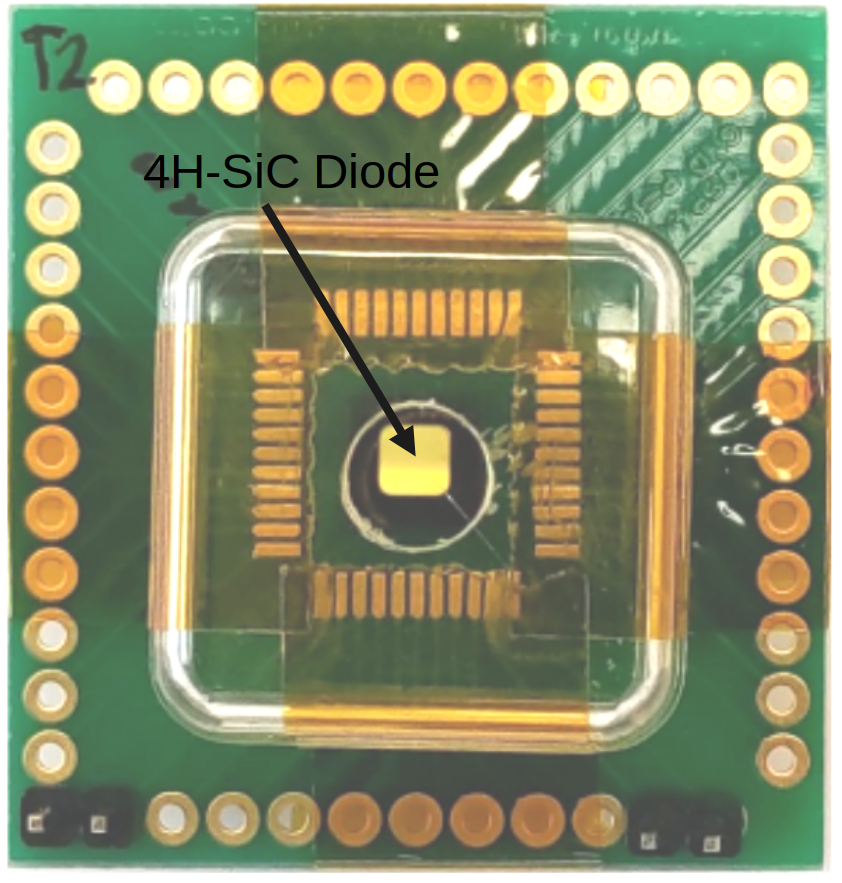}
    \caption[]{\textit{Left:} Cross section (not to scale) of the used 4H-SiC diode designed by IBS. It features a two-stage epitaxial p-type anode on top of a \SI{110}{\micro\meter} n-type epitaxial bulk on a 5.8 $\times$ \SI{5.8}{\square\milli\meter} die with a pad size of 2.9 $\times$ \SI{2.9}{\square\mm}. \textit{Right:} Picture of the 4H-SiC diode mounted on a carrier PCB.}
    \label{fig:SchemticSiC}
\end{figure}

The 4H-SiC diode was designed and manufactured by Ion Beam Services (IBS) \cite{IBS}. It is fabricated on a \SI{110}{\micro\meter} epitaxial substrate with a two-stage epitaxial p-type anode on top. No breakdown is observed up to reverse bias voltages of \SI{1000}{\volt}. The diode is glued and wire-bonded onto a carrier PCB. A plastic cover with a hole protects the wire bonds and facilitates tests with radioactive sources. A schematic cross section of the 4H-SiC diode as well as a picture of the mounted diode are shown in Figure \ref{fig:SchemticSiC}. The used silicon diode is the commercially available BPW34F diode with a sensitive area of \SI{7.02}{\square\mm} \cite{OSRAM_BPW34F}. 
\subsection{Setup}
The diodes are placed in a custom-made DUT box and their bias voltage is provided by a Keithley 2410 source measurement unit through the charge sensitive amplifier (CoolFET by Amptek). The readout chain consists of the amplifier that is directly connected to a multi-channel analyzer (Px5 by Amptek), enabling the measurement of different X-ray spectra.

The SiC and Si diodes are operated at reverse bias voltages of \SI{800}{\volt} and \SI{80}{\volt}, respectively. This ensures significant depletion for both diodes. We assume full charge collection efficiency for both devices, since the structures are single-pad diodes and the full-energy peak is used to derive the $\omega_{SiC}$ result. The \SI{110}{\micro\meter} epitaxial layer of the 4H-SiC diode requires a bias voltage larger than \SI{2000}{\volt} for full depletion, which was not realized in this setup.

An $^{241}$Am and a $^{109}$Cd radioactive point source as well as a variable X-ray source, based on X-ray fluorescence, are used. Both point sources emit high energy photons marked as $\gamma$ in Table \ref{tab:Eweight}, while Cadmium additionally emits X-rays via electron capture. With the variable source, photons from a radioactive decay excite the atoms of a selectable target. The used targets are $^{158}$Tb and $^{133}$Ba, which emit X-rays after excitation with characteristic energies $K_{\alpha1}$,  $K_{\alpha2}$, $K_{\beta3}$, $K_{\beta1}$, $K_{\beta2}$ shown in the upper part of Table \ref{tab:Eweight}.

\section{Measurement results}\label{sec:Result}
An energy calibration can be performed by fitting a Gaussian function to each gamma peak in the observed spectrum. Since the individual X-ray lines of Terbium, Barium and Cadmium can not be distinguished in the silicon nor in the 4H-SiC diode, an intensity-weighted mean is calculated ($K_{\alpha}$, $K_{\beta}$ and $K_{\alpha+\beta}$). These X-ray energies as well as the high-energy $\gamma$ used for the energy calibration are given in the lower part of Table \ref{tab:Eweight}. 
\begin{table}[h]
     \centering
     \small

\begin{tabular}{c|cc|cc}
\hline
& \multicolumn{2}{c|}{Variable source}
& \multicolumn{2}{c}{Point sources} \\
\cline{2-5}
X-ray & $^{158}$Tb & $^{133}$Ba & $^{109}$Cd & $^{241}$Am \\
\hline
\multicolumn{5}{c}{\textbf{Characteristic X-ray energies (keV)}} \\
\hline
$K_{\alpha 2}$ & 43.744 & 31.817 & 21.990 & -- \\
$K_{\alpha 1}$ & 44.482 & 32.194 & 22.163 & -- \\
$K_{\beta 3}$  & 50.229 & 36.304 & 24.192 & -- \\
$K_{\beta 1}$  & 50.382 & 36.378 & 24.943 & -- \\
$K_{\beta 2}$  & 51.698 & 37.257 & 25.455 & -- \\
\hline
\multicolumn{5}{c}{\textbf{Energies used for calibration (keV)}} \\
\hline
$K_{\alpha}$       & 44.22 & 32.06 & -- & -- \\
$K_{\beta}$        & 50.59 & 36.51 & -- & -- \\
$K_{\alpha+\beta}$ & 45.44 & 32.87 & 22.53 & -- \\
$\gamma$           & -- & -- & 88.03 & 59.54 \\
\hline
\end{tabular}

     \caption[Weighted energy means]{The characteristic X-ray and $\gamma$ lines of the used radioactive sources. The intensity-weighted means of the $K_{\alpha}$ and $K_{\beta}$ lines used for the calibration are additionally shown. The characteristic X-ray energies of $^{158}$Tb and $^{133}$Ba are obtained from \cite{x-ray}. The decay energies of $^{109}$Cd and $^{241}$Am can be found in \cite{Sr90}. The intensity-weighted means for $K_{\alpha}$ and $K_{\beta}$ are calculated with their respective intensities for Terbium, Barium and Cadmium. $K_{\alpha+\beta}$ represents the intensity-weighted mean of both $K_{\alpha}$ and $K_{\beta}$.}
     \label{tab:Eweight}
\end{table}

\begin{figure}[h]
    \centering
    \includegraphics[width=0.8\linewidth]{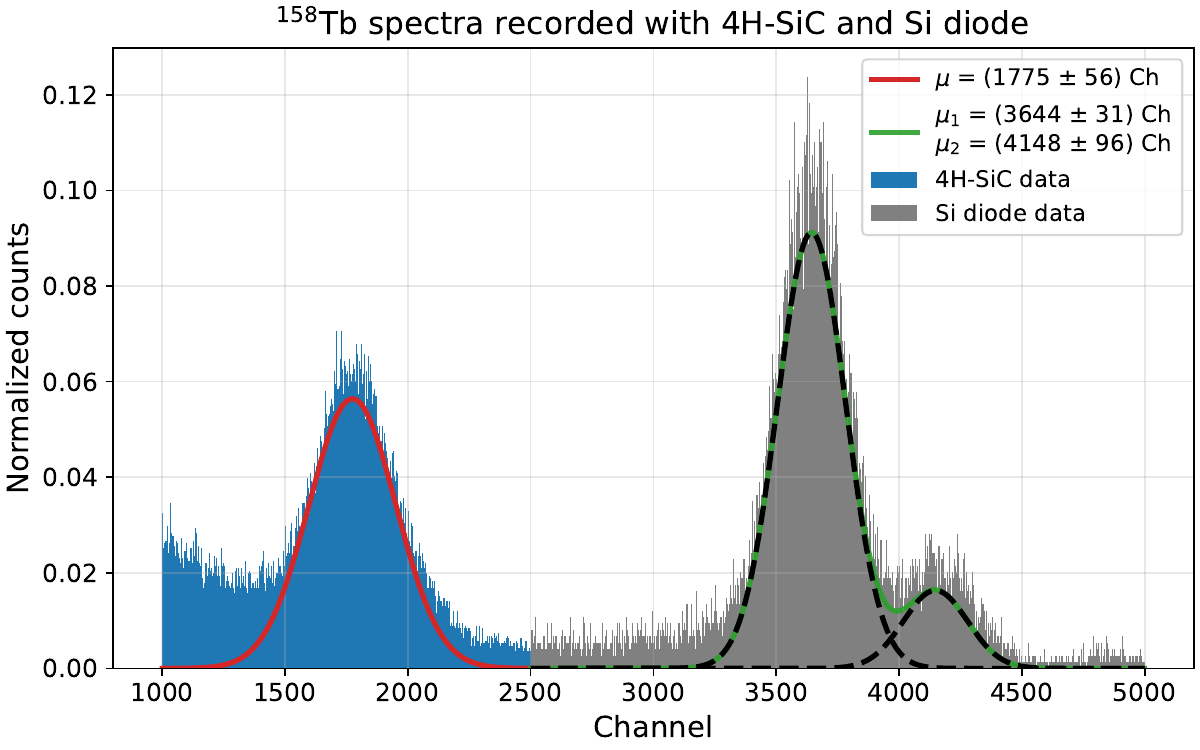}
    \caption{The results of fits of Gaussian functions to the peaks of the spectra for $^{158}$Tb recorded with the 4H-SiC and the Si diode. A larger amplitude and better energy resolution are observed for the silicon diode. Consequently, the weighted mean energies of the $K_{\alpha}$ and $K_{\beta}$ lines are used separately for the energy calibration of the silicon diode, whereas a single weighted mean $K_{\alpha+\beta}$, combining both lines, is used for the 4H-SiC detector.}
    \label{fig:spectra}
\end{figure}

\begin{figure}
    \centering
    \includegraphics[width=0.8\textwidth]{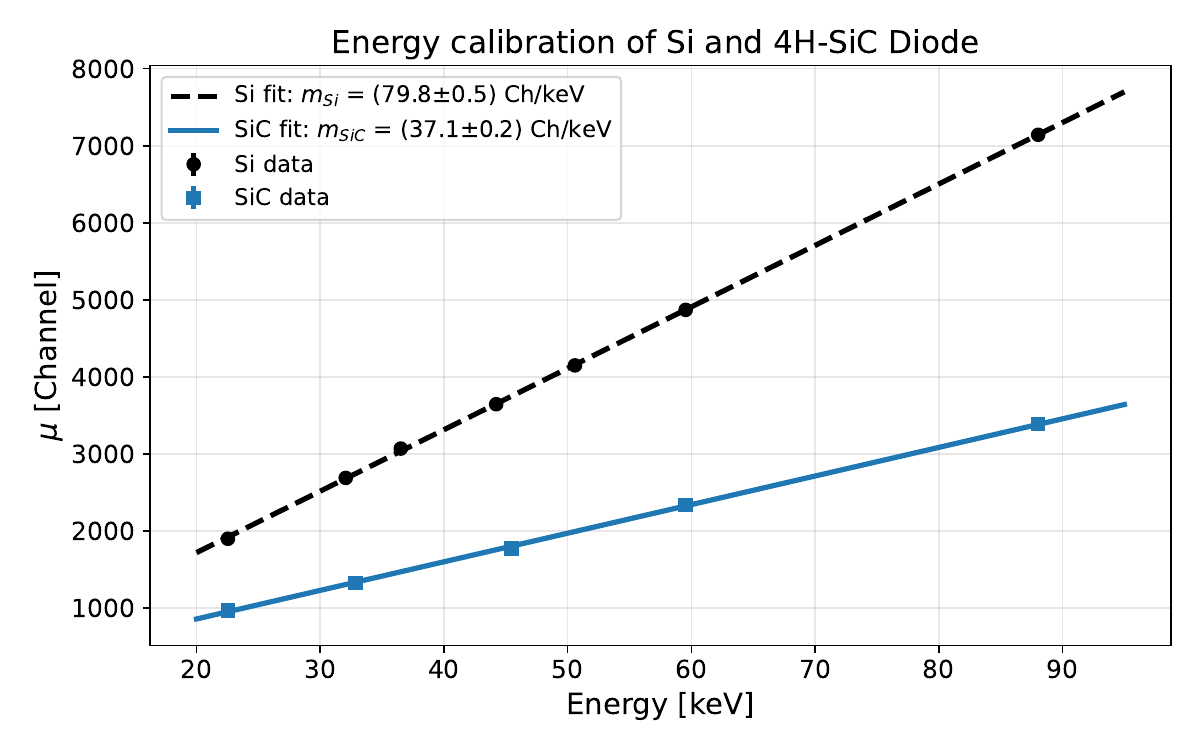}
    \caption[]{Energy calibration required to determine $\omega_{SiC}$. For the SiC diode, fewer calibration points are available due to its worse energy resolution compared to silicon. The uncertainty bars are too small to be visible.}
    \label{fig:Result}
\end{figure}

Since silicon has a smaller electron-hole pair creation energy, its energy resolution is better than for 4H-SiC. With the Si diode the $K_{\alpha}$ and $K_{\beta}$ peaks of Barium and Terbium can be distinguished, while for the 4H-SiC diode they can not. A weighted mean $K_{\alpha+\beta}$ of both $K_{\alpha}$ and $K_{\beta}$ lines is therefore used for SiC, which leads to fewer available data points for the energy calibration. Due to the smaller electron-hole pair creation energy of silicon, the occurring amplitudes and therefore the fitted channel numbers are higher for silicon compared to 4H-SiC. An exemplary spectrum for Terbium is shown in Figure \ref{fig:spectra}, highlighting the differences between silicon and SiC.
Poisson errors are assumed for every channel entry. A noise term $\sigma_{\text{FET}}$ for the CoolFET amplifier (\SI{284.5}{\electronvolt} \cite{FET}) is added to the fit uncertainty, $\sigma_{\mu}$, of each peak position: 
\begin{equation}
    \sigma^2=\sigma^2_{\mu}+\sigma^2_{\text{FET}}.
\end{equation}
The obtained peak positions are plotted against the respective photon energy. Linear fits are used to determine the slopes $m_i$ needed for the energy calibration and shown in Figure \ref{fig:Result}.
With the determined slopes for the SiC and Si diodes, $\omega_{SiC}$ can be calculated according to:
\begin{equation}\label{eq:w_sic}
    \omega_{SiC}=\omega_{Si}\cdot\frac{\mu_{Si}}{\mu_{SiC}}=\omega_{Si}\cdot\frac{m_{Si}}{m_{SiC}}=(7.87\pm0.09)\,\text{eV}
\end{equation}
The uncertainty is governed by the uncertainties of the linear fits and, therefore, by the number of available data points. Each data point includes a fit uncertainty, $\sigma_{\mu}$, and a noise term of the amplifier, $\sigma_{\text{FET}}$. In addition, the uncertainty of $\omega_{Si} = \SI{3.66+-0.03}{\eV}$ is included in the total uncertainty for $\omega_{SiC}$ through Gaussian error propagation using Equation \ref{eq:w_sic}, in contrast to the previous publications.
\subsection{Comparison to previously published results}
A summary of other previously published results that also compare signals in silicon and 4H-SiC, is shown in Table \ref{tab:compare}. Compared to our measurement, the X-ray-based measurements published in \cite{Bertuccio2003} and \cite{Phlips2005} focus their investigations solely on $^{241}$Am as a radioactive source, do not include an uncertainty on $\omega_{Si}$ and do not quote one for $\omega_{SiC}$. This is also partly the case for the previously reported $\alpha$-measurements \cite{Lebedev2004} \cite{Ivanov2004}. For these measurements, it is furthermore essential that the active sensor thickness $d$ is larger than the range of the used $\alpha$ particles. Otherwise, it must be taken into account that the particles do not deposit their entire energy in the sensor. In publication \cite{Lebedev2004}, a 4H-SiC detector with an active sensor thickness of \SI{10}{\micro\meter} is employed. The range of the $\alpha$ particles of approximately \SI{20}{\micro\meter} significantly exceeds the active sensor thickness. This is not the case for the other $\alpha$-based measurements and may therefore contribute to the large discrepancy between the reported values of $\omega_{SiC}$. 
\begin{table}[h]
     \centering
     \small
\begin{tabular}{c|c|cc|cc|c|c}
\hline
Source & $d$ ($\mu$m) & $\omega_{\mathrm{Si}}$ (eV) & $\Delta\omega_{\mathrm{Si}}$ (eV)
& $\omega_{\mathrm{SiC}}$ (eV) & $\Delta\omega_{\mathrm{SiC}}$ (eV) & Year & Ref. \\
\hline
X-ray / $\gamma$ & -- & 3.66 & 0.03 & 7.87 & 0.09 & 2026 & This work \\
X-ray ($^{241}$Am) & -- & 3.60 & -- & 7.60 & -- & 2005 & \cite{Phlips2005} \\
X-ray ($^{241}$Am) & -- & 3.67 & -- & 7.80 & -- & 2003 & \cite{Bertuccio2003} \\
\hline
$\alpha$ & 50 & 3.62 & -- & 7.83 & 0.02 & 2023 & \cite{wien} \\
$\alpha$ & 35 & 3.62 & 0.02 & 7.78 & 0.05 & 2005 & \cite{LoGiudice2005} \\
$\alpha$ & 10 & 3.62 & -- & 8.60 & -- & 2004 & \cite{Lebedev2004} \\
$\alpha$ & 20 & 3.62 & -- & 7.71 & -- & 2004 & \cite{Ivanov2004} \\

\hline
Proton & 35 & 3.64 & 0.02 & 7.79 & 0.09 & 2005 & \cite{LoGiudice2005} \\
\hline
\end{tabular}

\caption{Previously published values of the electron-hole pair creation energy in 4H-SiC ($\omega_{\mathrm{SiC}}$), determined by comparison with silicon, as done in this work. The radiation sources, active sensor thicknesses $d$, and corresponding silicon electron-hole pair creation energies $\omega_{\mathrm{Si}}$ used for the comparison are also listed. The quantities $\Delta\omega_{\mathrm{Si}}$ and $\Delta\omega_{\mathrm{SiC}}$ denote the uncertainties of $\omega_{si}$ and $\omega_{SiC}$, respectively, as reported in the original publications. Dashes indicate that no uncertainty was specified.}\label{tab:compare}
\end{table}

\begin{figure}[h]
    \centering
    \includegraphics[width=0.5\linewidth]{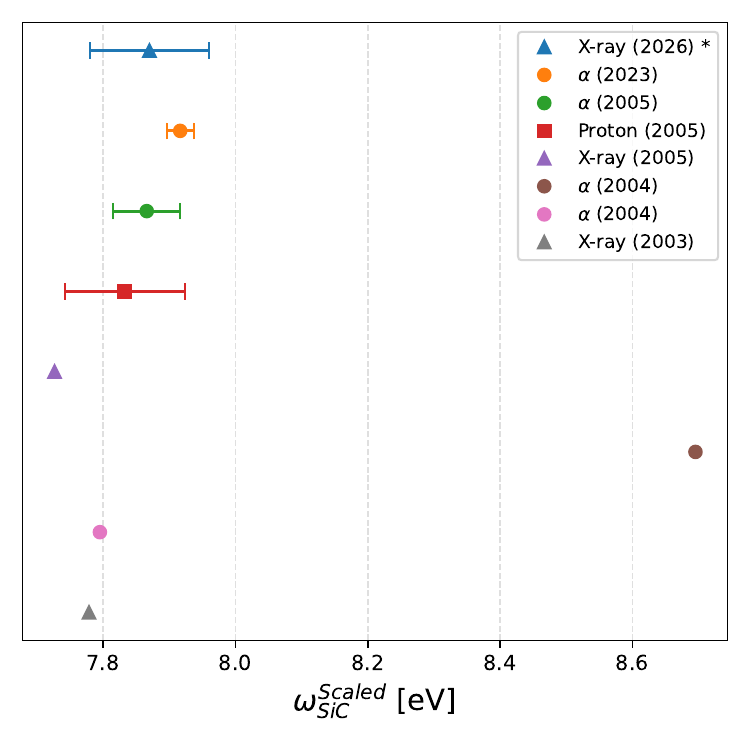}
\caption{Rescaled values of previously published electron-hole pair creation energies in 4H-SiC ($\omega^{Scaled}_{SiC}$), normalized to the $\omega_{Si}$ value used in this work. In chronological order: \cite{Bertuccio2003}\cite{Ivanov2004}\cite{Lebedev2004}\cite{Phlips2005}\cite{LoGiudice2005}\cite{LoGiudice2005}\cite{wien}. The result marked by a "*" represents the value obtained from the measurement presented in this paper. For the results shown without uncertainty bar, no uncertainty was quoted in the original publications.}\label{fig:compare}
\end{figure}

To account for differences in the reported $\omega_{\mathrm{SiC}}$ results arising from variations in the used value of $\omega_{\mathrm{Si}}$, the published values are rescaled to the silicon electron-hole pair creation energy used in this work, according to Equation \ref{eq:scale}. 
\begin{equation}\label{eq:scale}
    \omega^{scaled}_{SiC} = \omega_{SiC}\times\frac{3.66\,\mathrm{eV}}{\omega_{si}} 
\end{equation}
Figure \ref{fig:compare} displays the comparison of previous $\omega_{SiC}$ determinations after rescaling. After rescaling the deviation of \cite{Lebedev2004} from the other results still persists. Within the reported uncertainties, our result is consistent with the previously published values of $\omega_{SiC}$. 
\section{Summary}
The electron-hole pair creation energy in 4H-SiC is determined to \SI{7.87 +- 0.09}{\electronvolt}. This value has been obtained from a comparison of the response for a Si and a 4H-SiC diode to X-ray and $\gamma$ radiation using $\omega_{Si} = \SI{3.66+-0.03}{\eV}$. This is the first X-ray-based determination of $\omega_{SiC}$ using several lines from four different X-ray and $\gamma$ sources ($^{158}$Tb, $^{133}$Ba, $^{109}$Cd, and $^{241}$Am) and including an estimate of the uncertainty. The uncertainty of our result includes the uncertainty from the fit to the full-energy peak, the noise term of the employed amplifier and the uncertainty on $\omega_{Si}$. A direct comparison of the precision of this result and the level of agreement with previous measurements using single X-ray sources are not possible, since no uncertainties are quoted for the previous results. For the comparison with other measurements, the $\omega_{SiC}$ results have been rescaled using the value of $\omega_{Si}$ adopted in our study. Our result is consistent with the rescaled $\omega_{SiC}$ results of all measurements that quote an uncertainty, including the $\alpha$-based measurements.

\acknowledgments

The authors thank Ion Beam Services for providing the silicon carbide diodes. Supported by the Deutsche Forschungsgemeinschaft (DFG, German Research Foundation) under Germany’s Excellence Strategy – Cluster of Excellence “Color meets Flavor”, EXC 3107 – Project-ID 533766364”.



\bibliographystyle{JHEP}
\bibliography{biblio.bib}






\end{document}